\documentclass[runningheads]{llncs}
\usepackage[T1]{fontenc}
\usepackage{graphicx}
\usepackage{amsmath,amsfonts,amssymb}
\usepackage{graphicx}
\usepackage{multirow}
\usepackage{lscape}
\usepackage{blindtext}
\usepackage[colorlinks=true, allcolors=black]{hyperref}
\usepackage{booktabs,caption}
\usepackage[flushleft]{threeparttable}
\usepackage{graphicx}
\usepackage{float}
\usepackage[margin=42mm]{geometry}                
\usepackage[utf8]{inputenc}                       
\usepackage{url}                                  
\usepackage{booktabs}                             
\usepackage{multirow}                             
\usepackage{amsfonts}                             
\usepackage{amsmath}                              
\usepackage{nicefrac}                             
\usepackage{microtype}
\usepackage{authblk}
\usepackage{booktabs}
\usepackage{graphicx}
\usepackage[numbers]{natbib}
\begin{document}

\title{Integrating Feature Understandability into Feature Selection for Human-Centred Tabular Explanations}
\titlerunning{Generating Human-Centred Tabular Explanations}

\author{
Nicola Rossberg\inst{1,2}\orcidID{0009-0005-3883-5833} \and
Bennett Kleinberg\inst{4}\orcidID{0000-0003-1658-9086} \and
Barry O'Sullivan\inst{1,2,3}\orcidID{0000-0002-0090-2085} \and
Luca Longo\inst{1,2,3}\orcidID{0000-0002-2718-5426} \and
Andrea Visentin\inst{1,2,3}\orcidID{0000-0003-3702-4826}
}

\authorrunning{N. Rossberg et al.}

\institute{
School of Computer Science and Information Technology,
University College Cork, Ireland
\and
Centre for Research Training in Artificial Intelligence,
University College Cork, Ireland\\
\email{n.rossberg@cs.ucc.ie}
\and
Insight Centre for Data Analytics,
University College Cork, Ireland
\and
Department of Methodology and Statistics,
Tilburg University}

\maketitle

\begin{abstract}

Existing human-centred XAI research predominantly evaluates explanations after they have been generated. However, this fails to provide actionable insights into improving accessibility. This study incorporates human-based feature understandability into feature selection to improve the generation of human-centred explanations. Natural language explanations for machine learning models trained on tabular data often cite influential features in the input data to explain a given decision. As a consequence, users' ability to understand the produced explanation hinges on their understanding of the cited features. To quantify this natural-language feature understandability, Rossberg et al. \cite{rossberg2025feature} proposed the Feature Understandability Scale, which assigns numerical `understandability scores' to features based on user feedback. To evaluate the degree to which these scores can improve machine learning explanation quality, this study collects understandability scores across two datasets, proposes an understandability-aware feature selection algorithm to select features that are both understandable and informative, and trains a random forest based on these feature subsets. The SHAP-based explanations generated by this model are evaluated using user feedback, and changes in explanation quality are assessed. The individual evaluation of the explanations showed that while the understandability-based explanations scored higher on average, the difference was not statistically significant. In contrast, the comparative evaluation found that participants preferred these explanations significantly more frequently than expected at the chance level. It is concluded that the integration of understandability can yield more understandable explanations at minimal cost to classification performance, although changes to quality require elucidation in larger cohorts. 

\end{abstract}

\section{Introduction}

In recent years, the use of machine learning has become increasingly pervasive across a wide range of domains, including impactful areas such as medicine and agriculture \cite{phillips2012ai, pallathadka2023impact, bao2023revolutionizing, sharifani2023machine}. However, while black-box models are often powerful in predictive performance, they may lack transparency, which is essential, especially in critical fields such as medicine and finance \cite{Raufi2024}. 
The ability to audit machine learning systems is crucial to prevent the deployment of inequitable models that result from biased data or poor training. To promote the auditability of black-box models, a wide range of explainable AI (XAI) techniques has been proposed \citep{wang2024roadmap}. 
Through the implementation of such techniques, unstable decisions, especially in situations of uncertainty \cite{RizzoL18Explainability}, or those made on the basis of irrelevant data, can be identified and potentially mitigated, thereby supporting the legal compliance and safety of machine learning models \citep{longo2024explainable}.

While the implementation of XAI techniques is thus important to build user trust and prevent model bias, explaining is a social process and thus requires consideration of the end-user \cite{miller2019explanation}. Current techniques to measure explanation quality hinge on post-hoc evaluation, building on user feedback or algorithmic computations to quantify the accessibility of the explanation \cite{sovrano2023objective, vilone2023development, holzinger2020measuring}. While this provides insight into the quality of the explanation, it does not show which aspects of the explanation need to be changed to improve the usability. For textual explanations, in which the model's inferences are converted and expressed in natural language \citep{vilone2021classification}, references to input features are one actionable lever to improve explanation quality, with user understanding hinging on their comprehension of the features included in the explanation. To thus create an actionable explanation evaluation, it is necessary to quantify understanding of the feature, rather than the explanation level. 

\begin{figure}
    \centering
    \includegraphics[width=0.8\linewidth]{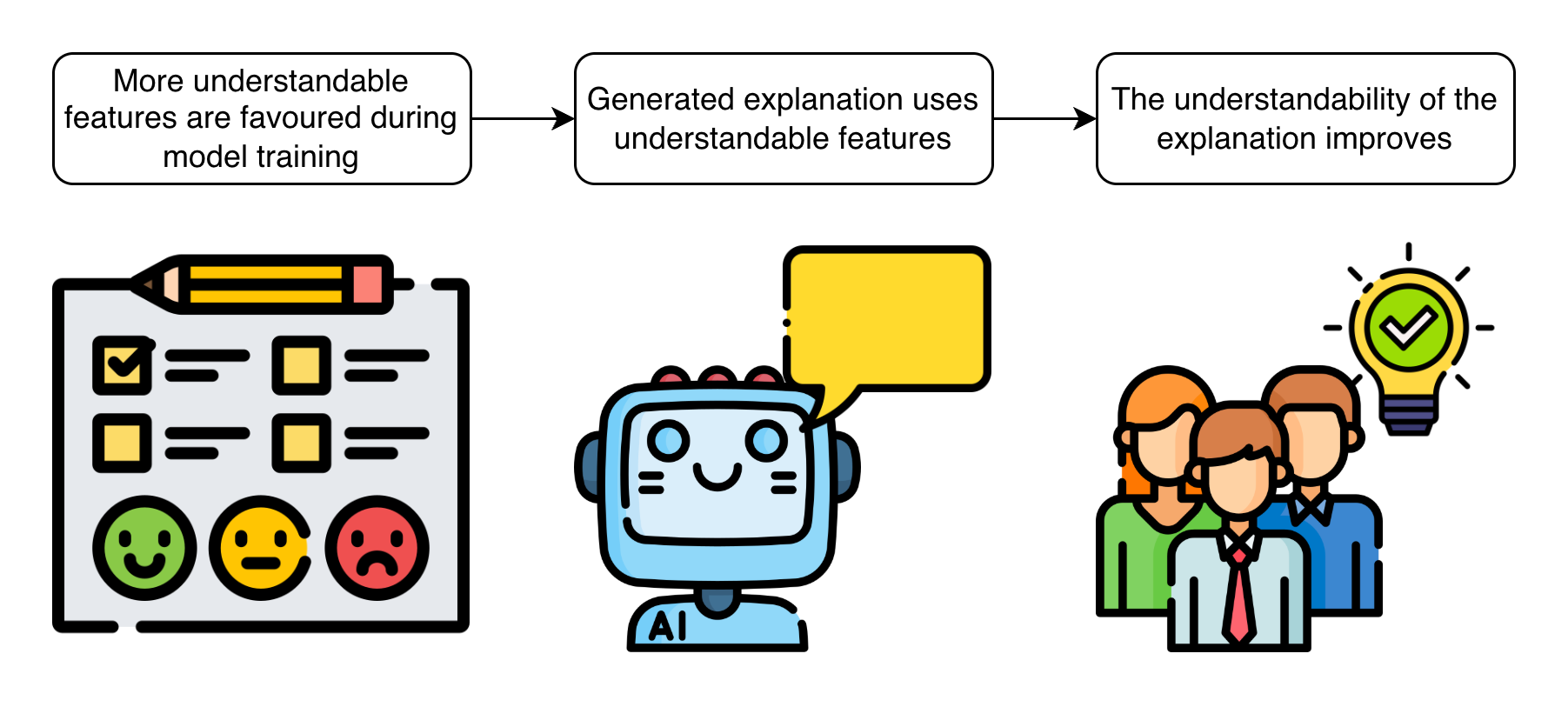}
    \caption{Graphical depiction of the theoretical basis on which the FUS implementation is expected to increase the understandability of explanations for tabular data.}
    \label{fig: Benefits FUS}
\end{figure}

To this end, Rossberg et al. \cite{rossberg2025feature} proposed the \textit{Feature Understandability Scale} (FUS), which measures feature understandability across various dimensions using user input. An example of the scale's implementation is provided in Table \ref{tab: example} in the Appendix. In the final use case, the scale is distributed to a representative sample of the audience receiving the final explanation. The understandability scores (US) assigned to each feature are computed by taking the average user rating. The US can then be incorporated into the machine learning pipeline through various modalities, including cost-sensitive feature selection or a Pareto frontier. By favouring more understandable features during model training, the explanations produced by the machine learning model are expected to be more accessible, thereby increasing their utility. This is illustrated in Figure \ref{fig: Benefits FUS}

This study aims to fulfil three successive objectives, which are outlined below:

\begin{enumerate}
    \item Collect a representative sample of understandability ratings across two datasets and analyse their distributions;
    \item Propose an understandability-aware feature selection mechanism, and train a machine learning model on the basis of the selected feature subsets;
    \item Evaluate the quality of the explanations generated on the basis of the trained machine learning model. 
\end{enumerate}

The rest of this report is structured as follows. Section \ref{sec: LitReview} provides an overview of the relevant literature pertaining to XAI, cost-sensitive feature selection and the explainability-accuracy trade-off. Section \ref{sec: Methodology} describes the steps taken to collect understandability ratings, build the feature selection algorithm and evaluate the generated explanations. Section \ref{sec: Results and Discussion} then presents and analyses the results of each of the three steps of the study. Finally, Section \ref{sec: Conclusion} summarises and evaluates the results and Section \ref{sec: Limitations} details the limitations and future research directions of this study.

\section{Literature Review} \label{sec: LitReview}

\subsection{Explainable Artificial Intelligence}

While the pervasiveness of Machine Learning models has increased across a range of domains in recent years, the dangers associated with their black-box nature, including bias and shortcut learning, have been significant barriers to their widespread implementation \cite{bassi2024explanation, liu2025improving, simuni2024explainable, shah2023enhancing}. XAI techniques address these concerns by auditing decision processes and promoting transparency and trust \citep{wiratsin2025effectiveness, heider2026assessing}. A large number of XAI techniques have been proposed, including feature attribution \cite{scholbeck2023algorithm, koenen2024toward}, counterfactuals \cite{kuhl2023better,baron2023explainable,hollig2023semantic}, latent-space mechanistic interpretability \cite{Ahmed2022,zimmermann2023scale,kastner2024explaining,ahmed2025latent}, and natural language methods \cite{danilevsky2021explainability,zini2022explainability, Vojtech2026}. A full review and taxonomy of XAI methods can be found at Vilone and Longo \cite{vilone2021notions}.

\subsection{Interpretability by Design}
One important aspect considered in recent works is the construction of models that are interpretable by design, integrating transparency into the model's architecture and training process \cite{rudin2019stop, gao2026towards}. Examples of such models may include learning of predictors through interpretable queries \cite{chattopadhyay2022interpretable}, the inclusion of human judgment in the model design process \cite{lage2018human}, and the integration of human feature evaluations \cite{zytek2022need, veeramachaneni2014towards}. While interpretability by design promotes faithful explanations for machine learning systems, it has been argued that it may lead to lower performance, as transparent algorithms lack the necessary complexity \cite{assis2023explainable, hunsicker2025investigating, shuvra2024explainability, kabir2025review, assis2025performance}. However, while this is theoretically sound, it is seldom observed in practice, and it is therefore argued that more transparent models should be favoured by default \cite{bell2022s, dziugaite2020enforcing, carrington2018measures}.

\subsection{Cost-Sensitive Feature Selection}
Cost-sensitive feature selection incorporates cost into the feature selection framework, where cost is either the test cost or the misclassification cost of including the feature \cite{benitez2019cost, liu2017cost, bian2016efficient}. Test costs refer to the cost of feature acquisition, for example, through medical tests, where a more expensive test may yield a more accurate result \cite{jiang2019wrapper}. Costly misclassifications are domain-specific and user-specified. An example is the misclassification of cancerous cells as benign, which is more costly to public health than misclassifying healthy cells as cancerous, and features may be selected to promote false positives over false negatives \cite{zhou2016cost}.

In the literature, three types of cost-sensitive feature selection methods are distinguished: filter methods, wrapper methods and embedded methods \cite{zhao2019cost}. This work focuses on filter methods, as they are the most transparent and straightforward to implement. Filter methods pre-select features before model training, using feature cost as a scoring function. Features are then selected based on a user-defined threshold or feature number, selecting only the N cheapest features. Filter methods are advantageous as they are computationally cheap and can be applied to high-dimensional datasets \cite{bach2017cost}. 

This work differs from the traditional literature on cost-sensitive feature selection in that feature costs are user-defined rather than error- or test-based, and success is evaluated based on explanation quality rather than misclassification patterns. Where in the traditional literature the cost of a feature is defined either by its contribution to a costly misclassification or through the cost of its acquisition, this work defines a costly feature as one that is not understandable and thus contributes to ineffective explanations \cite{liu2017cost, zhou2016cost, bian2016efficient}. Unlike traditional literature, success is assessed by the final understandability of explanations and the accuracy of the machine learning model. 



\subsection{SHapley Additive exPlanations}

SHapley Additive exPlanations (SHAP) is based on cooperative game theory \cite{shapley1953value} and was among the earliest XAI techniques to provide both global (whole-model) and local (instance-specific) explanations. The implementation of SHAP values to measure feature importance was first proposed by \cite{lipovetsky2001analysis, sahakyan2021explainable} and later expanded for applications to Neural Networks \cite{keinan2004fair}, feature selection \cite{cohen2007feature} and outcome explanation \cite{vstrumbelj2009explaining}. However, the widespread use of SHAP began only when \cite{lundberg2017unified} published their unified framework, which combined previous approaches to create an XAI method that provides both local and global explanations, with model-agnostic and model-specific modalities \cite{sahakyan2021explainable}.

\subsection{Explanation Evaluation}

The evaluation of explanation quality has been investigated in past literature, and several different approaches have been proposed \cite{zhou2021evaluating, balog2020measuring, yang2019evaluating, mohseni2021quantitative, domnich2025predicting, dieber2022novel}. The properties of a good explanation are user and context-dependent. In general, it is important that the explanation be on the right level and include sufficient detail for the targeted user group \cite{kaplan2024unified}. The length of the explanation should also be tailored to the usage scenario; where a lay-user may have less time to read an explanation, a practitioner or data scientist may be willing to spend the time parsing a longer explanation \cite{chen2022makes}. When decomposing explanations to their purpose, Halpern et al. \cite{halpern2005causes} propose that a good explanation provides an answer to a 'why' question. In addition, Hoffman et al. \cite{hoffman2018metrics} state that a good explanation permits an a priori judgement regarding its quality and hence must be both precise and clear. While these metrics permit intermittent evaluation of explanations through qualitative analysis, the final assessment of an explanation should be based on user feedback. 

The collection of user-based evaluations of explanations should be reproducible and consistent. In the field of psychometrics, this is achieved through the development and validation of scales that reliably capture user attitudes \cite{carpenter2018ten}. Vilone and Longo \cite{vilone2023development} proposed a 12-item scale to measure the quality of machine-generated rule-based explanations, in which the user is queried about the clarity they have gained from the explanations. In a similar vein, Holzinger et al. \cite{holzinger2020measuring} proposed the system causability scale, which measures the quality of a human-AI interface or explanation process. This allows the user to assess the extent to which the communication method makes the provided explanations accessible.

\subsection{Synthesis and Literature Gaps}
This study aims to fill several gaps identified in the existing literature. To the best of the authors' knowledge, there are currently no studies integrating self-reported feature understanding into the machine learning training pipeline on an a priori basis. Second, the study expands on existing cost-sensitive feature selection methods by introducing a new approach to cost quantification and model evaluation. Finally, the study proposes two statistically rigorous explanation evaluation procedures, based on user preference and individual, scale-based evaluations.

\section{Method and Design} \label{sec: Methodology}

The methodology of this research is divided into three key steps. First, the FUS is applied across two datasets, and the US for each feature are computed. Second, two feature selection approaches are implemented in parallel: the first implements the understandability-sensitive feature selection, identifying features that are both informative and understandable, referred to as the co-optimised setting going forward. The second selects features solely based on informativeness without considering understandability, and is henceforth referred to as the informative setting. For each approach and dataset, a random forest model is implemented, optimised and evaluated. In the final step, SHAP is used to extract feature importance from the trained machine learning models, and a heuristic is applied to generate textual explanations. The explanations are evaluated through user feedback, and differences in explanation quality between the co-optimised and informative settings are analysed. The code for the complete analysis is available on GitHub\footnote{\url{github.com/ncrossberg/Explanation-Evaluation-Paper}.}

\subsection{Step 1 - Understandability Score Collection}

The FUS is applied to two public datasets to ensure that any changes in model explanation quality are not domain-specific. An example of the implementation of the FUS is shown in Table \ref{tab: example} in the Appendix, and the full scales can be found in \cite{rossberg2025feature}. The first dataset is the `Telco Customer Churn' dataset, which contains data that can be used to predict whether a customer will switch service providers given their family and usage history. 
The dataset is publicly available \footnote{\url{https://www.kaggle.com/datasets/blastchar/telco-customer-churn}}. This dataset will be referred to as the `telco dataset' going forward. The second dataset is the `Body Signals of Smoking' dataset and contains data useful for the prediction of whether a patient is a smoker, given a range of physiological indicators. The dataset is publicly available \footnote{\url{https://www.kaggle.com/datasets/kukuroo3/body-signal-of-smoking}} and will be referred to as the `medical dataset' going forward. Before collecting the US, the datasets were cleaned by removing features with zero variance. Furthermore, user ID variables were removed because they were unique patient identifiers and lacked predictive relevance. No other features were removed during preprocessing. The final telco dataset consisted of 19 features and 7043 instances, and the medical dataset consisted of 24 features and 55,692 instances. It is important to note that while diverse datasets are chosen to promote theoretical generalisability, practical transferability of findings to other domains cannot be guaranteed. 

US of the remaining features after preprocessing were collected via a Qualtrics\footnote{Qualtrics is a survey-building and distribution website \url{https://www.qualtrics.com/}}-based survey, distributed through Prolific\footnote{Prolific connects researchers with participants, enabling large-scale data collection: \url{https://www.prolific.com/}}. All data collection was approved by the University College Cork Ethics Committee for Social Research under Log Number 2025-231A4. Participants had to be over 18 and be nationals of, and reside in, the United States of America, the United Kingdom, or Ireland to be eligible. All participants had to provide informed consent before commencing the study.

Participants were randomly assigned to one dataset and asked to rate half of the features using the FUS. Each participant evaluated a total of 10 features (telco dataset) or 12 (medical dataset). Participants were asked to rate only half the features to limit workload and avoid respondent fatigue. Features were presented in a pseudo-random order and counterbalanced to ensure that each feature received an approximately equal number of ratings. Feature-wise attention checks were implemented, and features with failed checks were removed. Finally, the mean US were computed for each feature. The features and respective mean and total US can be found on GitHub.

\subsection{Step 2 - Feature Selection and Machine Learning Model Implementation}

To ensure that the reported performance results of the final models were not due to the chosen train-test split, a repeated Monte Carlo sampling procedure was implemented prior to feature selection and machine learning model training to assess the equivalence of different data splits. The data were split into a stratified 70:30 train-test split across 50 different seeds to assess the distribution of model performance. The mean, standard deviation and normal-approximation confidence intervals were computed across the fifty splits. 

\begin{figure}[!ht]
    \centering
    \includegraphics[width=0.8\linewidth]{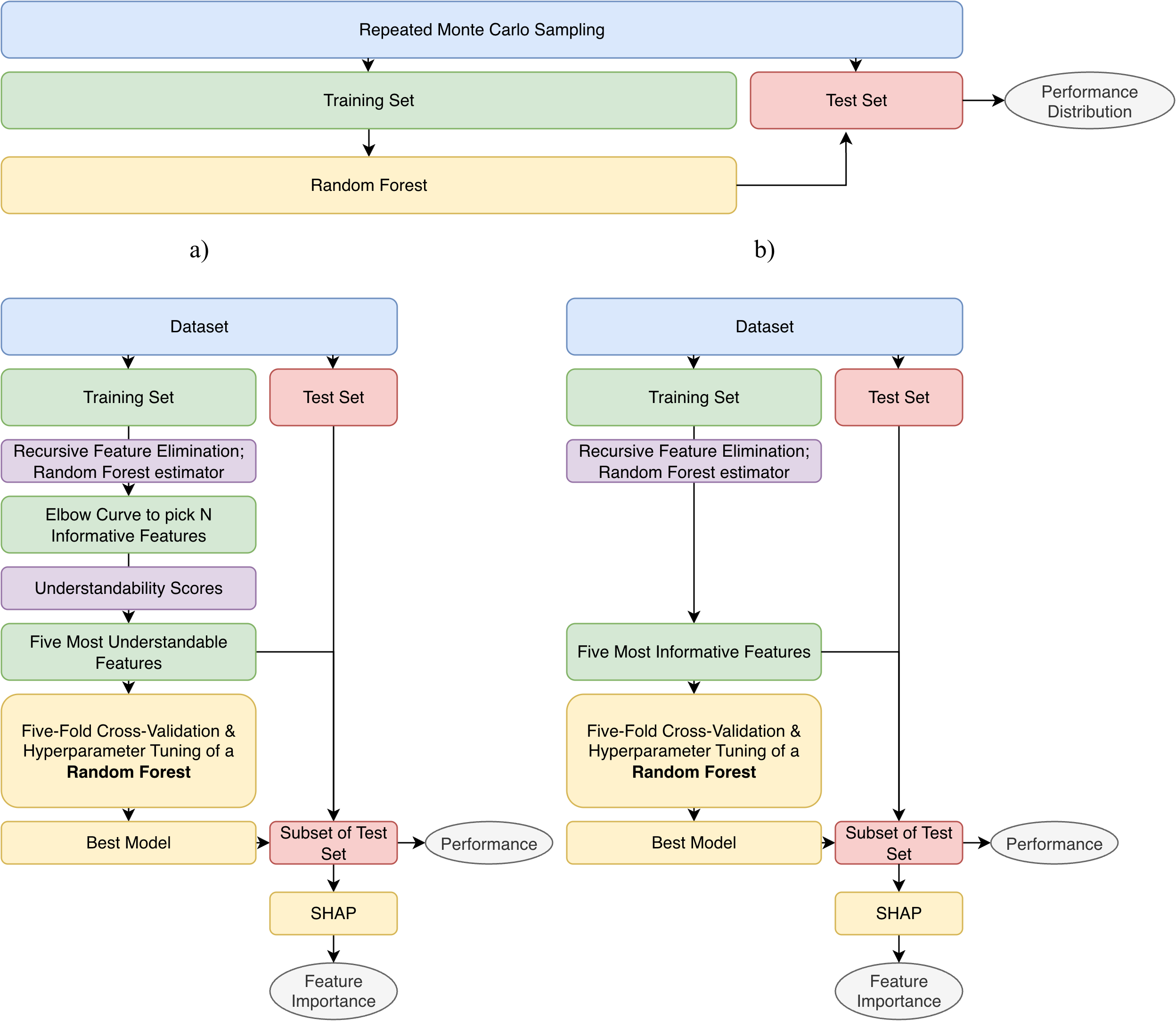}
    \caption{Visualisation of the Repeated Monte Carlo Sampling, a) co-optimisation and b) informative setting of the feature selection, and machine learning model training methodology.}
    \label{fig: Methodology_Schematic}
\end{figure}

Subsequently, the data were split into training and testing sets in a 70:30 ratio, and the subsequent procedures were implemented on the training set only. Figure \ref{fig: Methodology_Schematic} shows the full workflow of this step. 

\subsubsection{Feature Selection}
Two feature selection procedures were implemented in parallel, for the co-optimised and informative setting, respectively. Feature informativeness for both settings was computed through recursive feature elimination (RFE) with a random forest estimator. In the co-optimised setting, feature importance was plotted, and the top n features were identified using an elbow function. Of this subset of size n, the five most understandable features were retained as the final feature subset. In the informative setting, the top five most important features were selected directly from the fitted RFE model. A final subset of size five was selected to constrain explanation length and limit cognitive load, while still providing sufficient information to the user.

\subsubsection{Machine Learning Model Implementation}

After the feature subset generation, a random forest classifier was trained and optimised. A random forest was chosen as the model because it is less prone to overfitting than decision trees, can efficiently handle categorical data, and is well-suited to large datasets. Hyperparameter tuning was implemented using five-fold cross-validation on the training set, and the best-performing model was retrained on the entire training set and evaluated on the final test set. To generate the final explanations, ten instances were selected from the test set, and TreeSHAP \cite{lundberg2020local} was implemented to extract feature importance values.

\subsection{Step 3 - Explanation Evaluation}

Based on the feature importances extracted in step 2, the final textual explanations were generated using the heuristic outlined below:

\begin{enumerate}
    \item Compute instance i's feature contributions through Tree SHAP
    \item Partition features into two sets by contribution direction;\\
    - SUPPORT = features increasing the predicted outcome\\
    - DETRACT = features decreasing it
    \item Verbalise each feature:\\
    - numeric $\rightarrow$ compare value to its mean ("above/below average")
    \item Compose explanation:\\
    - State prediction, justified by SUPPORT features (in rank order)\\
    - If DETRACT is non-empty, append as a caveat clause
    \item Return explanation string.
\end{enumerate}

Explanations were corrected for grammar using the software "Grammarly" to ensure cohesion and legibility. The generated explanations were evaluated through user feedback to assess differences in quality between the co-optimised and informative settings. Qualtrics was used to implement the evaluation procedure, and participants were recruited via Prolific. Data was collected on the 15th of September 2026, and participants were remunerated monetarily in line with Irish minimum wage requirements. To be eligible, participants had to be over the age of 18 and nationals of and residing in Ireland, the United Kingdom or the United States. They also could not have previously completed a survey related to this research. All participants had to digitally sign an informed consent form prior to commencing the survey.\footnote{The informed consent form can be found on Github \url{github.com/ncrossberg/Explanation-Evaluation-Paper}}

\begin{figure}[!h]
    \centering
    \includegraphics[width=\linewidth]{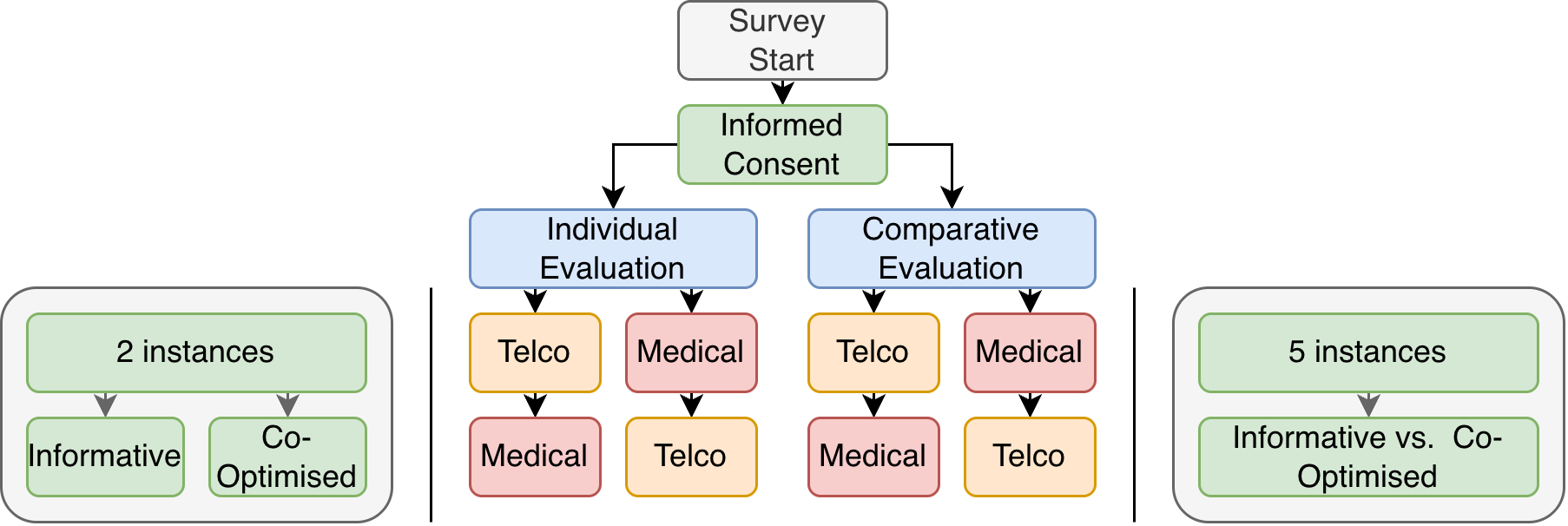}
    \caption{The Survey Flow of the evaluation procedure. Participants were randomly assigned to the comparative or individual evaluation, and the order of datasets was randomised for each condition. The individual setting used the explanation quality scale \cite{vilone2023development} to evaluate informative and co-optimised explanations separately, whereas the comparative setting prompted participants to select the more understandable explanation and provided prototypes of both co-optimised and informative explanations. Two or five instances were selected, respectively, and participants were asked to either use the scale to evaluate them (individual evaluation) or select the more understandable explanation (comparative evaluation).}
    \label{fig: Qualtrics Workflow}
\end{figure}

The explanation evaluation procedure was split into two distinct approaches: individual and comparative evaluation, the details of which are outlined below. A total of ten prototype explanations were selected for each dataset and explanation setting (40 explanations total). The explanations are expected to function as a prototype of the optimisation procedure, and a mixed effects model was implemented to assess the prototype-wise variation. Participants were randomly assigned to one of the evaluation procedures. The full survey workflow is outlined in Figure \ref{fig: Qualtrics Workflow}.

\subsubsection{Individual Evaluation}
The individual evaluation used an adapted version of the scale developed by  Vilone \& Longo \cite{vilone2023development}. The questionnaire was adapted by removing two items: "No rules in the explanatory method return surprising predictions" and "The explanatory method only includes the most relevant variables from the data". The items were excluded, as explanations in this study are not rule-based, and the training set consists of only five features, precluding participants from assessing whether these are the most relevant, as they do not have access to the full feature set prior to selection. The remaining items were adapted by replacing `explanatory method' with `explanation'. The final scale is shown in Table \ref{tab: Scale Structure} in the Appendix. 

During the evaluation, participants were presented with two pairs of two explanations (Co-optimised and informative) for each dataset and asked to rate the explanations using the scale. The order of datasets and explanation type was randomised and counterbalanced to prevent order effects. After data collection, responses were cleaned, relevant scale items were reverse-coded, and total scores were computed for each prototype and instance. The survey was structured as a repeated-measures, paired design, with participants evaluating multiple explanations across both datasets. To account for the random error introduced by participants, a mixed-effects model is implemented. Each dataset is analysed separately, with the explanation type used as a fixed effect to predict the total score, while accounting for variation between participants and prototypes by including them as random intercepts. 

\subsubsection{Comparative Evaluation}
The comparative evaluation setting presented participants with two explanations for the same instance, generated by the co-optimised and informative settings, respectively. Participants were asked to select the explanation they understood better. Participants were presented with five pairs of explanations for each dataset (twenty explanations total). Presentation order of datasets and explanations was randomised and counterbalanced to prevent order effects. 

The analysis of the comparative results was conducted in two steps. First, descriptive statistics were computed, providing the total and question-wise preference count for each dataset to assess the general trends in explanation understandability. Subsequently, a mixed-effects logistic regression model was implemented to assess the significance of the number of preferences while accounting for participant and prototype effects. Models were implemented separately for each dataset, and preference was used as the outcome variable, with a random intercept included for both the participant and prototype. The models were implemented using the `lme4' library \cite{bates2015fitting}.

\section{Results and Discussion} \label{sec: Results and Discussion}

\subsection{Feature Rating Collection}

\begin{figure}
    \centering
    \includegraphics[width=\linewidth]{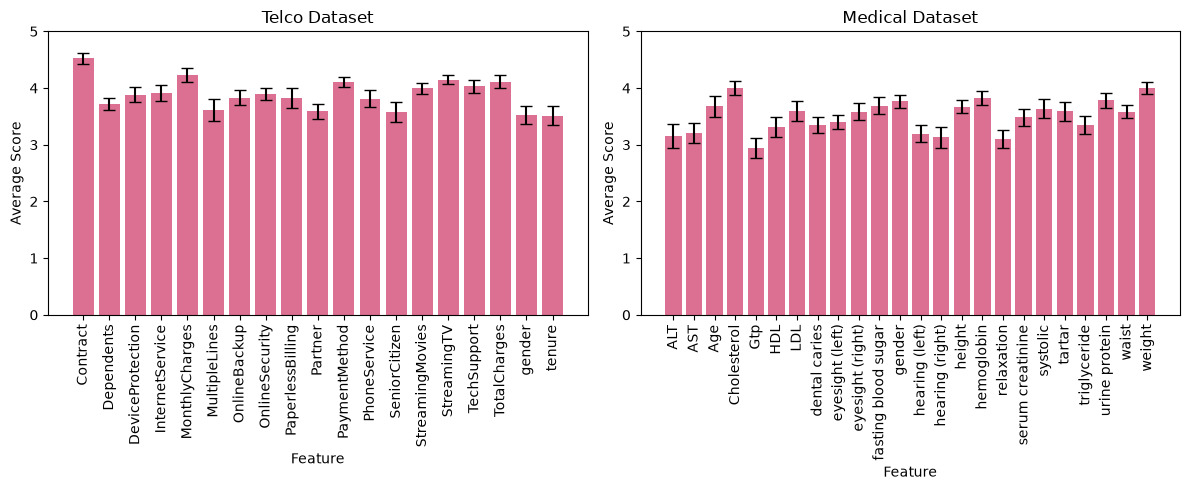}
    \caption{Distribution of US across the two datasets}
    \label{fig: Understandability Score Distribution}
\end{figure}

Feature ratings were collected on the 31st of August 2026. The average number of ratings per feature was 30.32 (minimum = 22, maximum = 37). Figure \ref{fig: Understandability Score Distribution} shows the distribution of mean US across both datasets, along with the standard errors for each feature, where higher scores indicate greater understandability. The range of US is limited across the two datasets (3.52-4.52 for the telco dataset and 2.94-4.00 for the medical dataset). Several hypotheses can be put forward for this behaviour. While it may be a genuine reflection of limited difficulty in the datasets, it may also be due to social desirability (users reporting inflated understanding to appear better) or overconfidence (users overestimating their own understanding). However, further testing is required to confirm these stipulations.

Interestingly, the telco dataset is considered more understandable on average, possibly due to the absence of specialised medical terms. However, cross-dataset comparisons should be treated with caution, since participants were only exposed to a single dataset, and they may have developed internal benchmarks based on the specific features they encountered. This lack of a shared baseline means that subjective interpretations of the rating scale were not standardised across the two groups, making direct comparisons of the US unreliable. Interestingly, features accessible at face value, such as `hearing (right)', were assigned comparatively low understandability, which may be attributable to the measurement units, and future interview studies may wish to investigate the factors leading to users' rating decisions. 

\subsection{Co-Optimisation}

\subsubsection{Monte Carlo Resampling}
The Monte Carlo resampling was conducted across fifty seeds, implementing a stratified 70:30 train-test split and training and evaluating a random forest on the full feature set for each split. The random forest model was implemented through Scikit-learn, and with the exception of balanced class weights, all default parameters were maintained, to isolate the effect of the split from that of hyperparameter tuning \cite{scikit-learn}. Table \ref{tab: Monte_Carlo_Distribution} reports the means, standard deviations and confidence intervals (computed through normal approximation) for the resampling procedure. 

This highlights that the variation in performance across seeds is limited, and the seed selection for the splitting procedures does not materially affect the reported results. Seed 42 was therefore maintained throughout both feature selection and modelling settings.

\begin{table}[]
\centering
\caption{Distributional Data of the Repeated Monte Carlo Sampling for both Datasets}
\label{tab: Monte_Carlo_Distribution}
\resizebox{0.8\columnwidth}{!}{%
\begin{tabular}{lcccc}
\hline
\multicolumn{5}{c}{\textbf{Telco Data}} \\ \hline
 & Mean & SD & Lower 95\% CI & Higher 95\% CI \\ \hline
Balanced Accuracy & 0.733 & 0.009 & 0.731 & 0.736 \\
F1 & 0.603 & 0.013 & 0.600 & 0.607 \\
Precision & 0.569 & 0.013 & 0.565 & 0.573 \\
Recall & 0.642 & 0.020 & 0.637 & 0.648 \\ \hline
\multicolumn{5}{c}{\textbf{Medical Data}} \\ \hline
Balanced Accuracy & 0.831 & 0.003 & 0.830 & 0.832 \\
F1 & 0.781 & 0.003 & 0.780 & 0.782 \\
Precision & 0.709 & 0.004 & 0.708 & 0.710 \\
Recall & 0.869 & 0.005 & 0.867 & 0.870 \\ \hline
\end{tabular}%
}
\end{table}

\subsubsection{Feature Selection}

The performance results of the co-optimisation procedure are shown in Table \ref{tab: co_op_results}. Notably, the performance of the random forest is stable across the co-optimised and informative settings. The largest observed difference is in the Telco dataset's recall, with the informative setting performing approximately 2\% better. Overall, the classifier's performance is stable across both settings, indicating that selecting features that are both understandable and informative has negligible effects on performance. This is furthermore supported by the changes in performance lying within or close to the standard deviation observed in the Monte Carlo sampling in the telco dataset, showing that the performance gap between settings is within the noise floor established prior. Importantly, the US of the selected features in the co-optimised setting is higher than that of the informative setting across both datasets. While the difference is rather small (0.20 and 0.25), it represents changes of 19.86\% and 22.63\% of the total US range in the two datasets, respectively. As such, the model increases the understandability by approximately 20\% of the observed range (5\% of the total range) without affecting classification performance.

\begin{table}[!ht]
\centering
\caption{Results of the Co-Optimisation Procedure for both datasets}
\label{tab: co_op_results}
\resizebox{0.8\columnwidth}{!}{%
\begin{tabular}{@{}lcccc@{}}
\toprule
 & \multicolumn{2}{c}{Telco Dataset} & \multicolumn{2}{c}{Medical Dataset} \\ \midrule
 & \multicolumn{1}{l}{Co-Optimised} & \multicolumn{1}{l}{Informative} & \multicolumn{1}{l}{Co-Optimised} & \multicolumn{1}{l}{Informative} \\ \midrule
Balanced Accuracy & 75.25 & 75.35& 75.54 & 75.40\\
F1 & 61.11 & 61.03 & 70.21 & 70.09\\
Precision & 49.29 & 48.48 & 55.64 & 55.40 \\
Recall & 80.39 & 82.35 & 95.11 & 95.37 \\
Mean US & 4.16 & 3.96 & 3.79 & 3.54 \\ \bottomrule
\end{tabular}
}
\end{table}

The selected features across the two datasets are presented in Table \ref{tab: feature_selection}. During initial feature selection in the co-optimised setting, nine features were retained for the telco dataset and eight for the medical dataset, using the elbow of the importance plot to determine the total cut-off. Of the five most understandable features retained, two overlap with the informative setting in the telco dataset and three in the medical dataset. At face value, the features selected by the co-optimised setting are more understandable, giving preference to variables which are more commonly encountered in everyday life, especially in the medical dataset.

\begin{table}[!ht]
\centering
\caption{Features selected for each dataset in the co-optimised and informative setting. The features are ranked by importance as determined by SHAP. Features selected by both settings are bolded.}
\label{tab: feature_selection}
\resizebox{0.8\columnwidth}{!}{%
\begin{tabular}{@{}clcl@{}}
\toprule
\multicolumn{2}{c}{Telco Dataset} & \multicolumn{2}{c}{Medical Dataset} \\ \midrule
Co-Optimised & \multicolumn{1}{c}{Informative} & Co-Optimised & \multicolumn{1}{c}{Informative} \\ \midrule
\multicolumn{1}{l}{\begin{tabular}[c]{@{}l@{}}Internet Service, \\ \textbf{Tech Support}, \\ Total Charges, \\ Monthly Charges, \\ Contract\end{tabular}} & \begin{tabular}[c]{@{}l@{}}Tenure, \\ Online Backup, \\ Online Security, \\ \textbf{Tech Support}, \\ Contract\end{tabular} & \multicolumn{1}{l}{\begin{tabular}[c]{@{}l@{}} \textbf{height(cm)}, \\ age, \\ \textbf{gender}, \\ \textbf{haemoglobin}, \\ weight(kg)\end{tabular}} & \begin{tabular}[c]{@{}l@{}}Gtp, \\ serum creatinine, \\ \textbf{height(cm)}, \\ \textbf{gender}, \\ \textbf{haemoglobin}\end{tabular} \\ \bottomrule
\end{tabular}%
}
\end{table}

\subsubsection{Explanation Generation}

A total of 40 prototype explanations are generated across the four categories (two settings and two datasets). An example of one of the explanations is presented below:

\begin{quote}
    The individual is predicted to change companies because their Contract is Month-to-month, their Internet Service is Fibre optic, their Tech Support is no, and their Monthly Charges of 96.05 are above average. The fact that their Total Charges of 1740.70 are less than average detracted from this prediction.
\end{quote}

\subsection{Evaluation} \label{subsec: Evaluation}

126 participants were recruited for the evaluation procedure. Participant ages ranged from 19 to 81, with a mean age of 42.76 (SD = 13.90); 61.1\% female. Each participant was assigned to either the individual or comparative evaluation and rated either 2 or 5 pairs of explanations per dataset, respectively. The responses were split by evaluation type and dataset. Before analysis, non-English speakers were removed from the data, removing three participants, and response times were then examined. Response times were processed separately for each evaluation, accounting for differences in task type and resulting response speed. A box-and-whiskers plot of response time was examined, and outliers were removed from the data, leading to the deletion of four participants (one participant from the individual and three from the comparative setting). The final datasets comprised 236 data points from 59 participants for the individual evaluation and 300 data points from 60 participants for the comparative setting, respectively.

\subsubsection{Individual Scale-based Analysis}

\begin{figure}
    \centering
    \includegraphics[width=0.7\linewidth]{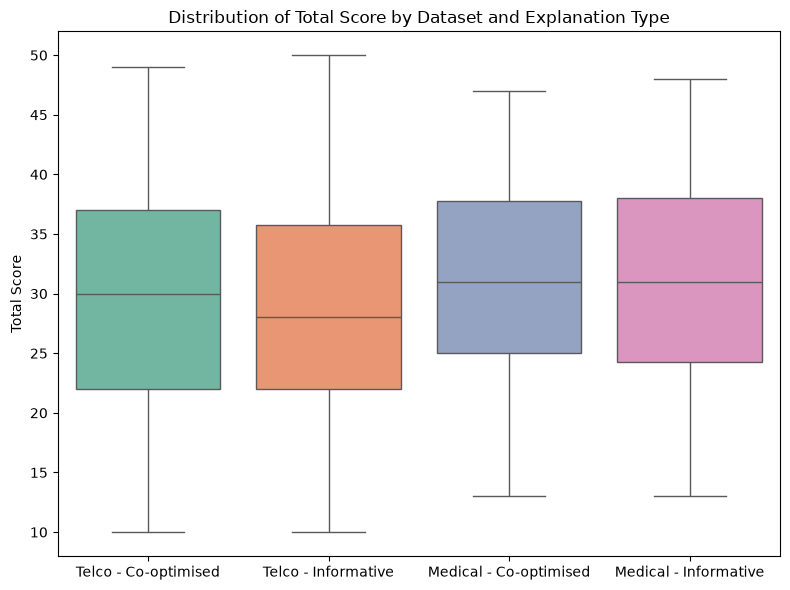}
    \caption{Distribution of Scale ratings across the two datasets and settings.}
    \label{fig: evaluation_distribution}
\end{figure}

Exploratory data analysis was then conducted. The distribution of total scores across the datasets and the two explanation types is shown in Figure \ref{fig: evaluation_distribution}. The co-optimised explanations score higher on average, although the difference is small ($\Delta_{Telco} = 1.03$ and $\Delta_{medical} = 0.08$). Given the repeated-measures design of the survey, a mixed-effects model is necessary to confirm whether the difference is statistically meaningful.

The two mixed effects models (one for each Dataset) were implemented using the `lme4' library in R \citep{bates2015fitting}. The p-values were estimated using Satterthwaite's method, and the confidence intervals are based on likelihood profile computations. The results for each model are presented in Table \ref{tab: MEM_results}. After the initial model specification, the assumptions of linearity, homoscedasticity, normality of residuals, normality of random effects and singularity were assessed and found to be satisfied.

The mixed-effects model predicted Total Score (range: 10-50) with random intercepts for Participants and Question prototype to account for repeated measures across participants (n=59) and questions (n=10) in both datasets. The model was implemented using the lme4 package in R \cite{bates2015fitting}. The results of both analyses are shown in Table \ref{tab: MEM_results}. Explanation type had no significant effect on total score, with the informative setting scoring lower than the co-optimised setting by an estimated 1.03 points (Telco) and 0.08 points (medical). While the quality change is in the expected direction, it is not sufficiently strong to reach statistical significance. Especially in the medical domain, the change is minimal and may be an artefact of sample fluctuations. While these results may genuinely reflect a lack of change in quality between explanations, they may also be an artefact of the study's small sample size, which lacks sufficient statistical power to reject the null hypothesis. Alternatively, the overlap in selected features, especially in the medical dataset, may lead to insufficient differentiation between explanations, consistent with the observed relative change in quality. Finally, the evaluation scale has not been psychometrically validated and hence may not permit the reliable evaluation of quality change. 

The Intraclass Correlation Coefficient (ICC) in Table \ref{tab: MEM_results} shows that of the variance not explained by explanation type, 55.4\% (telco) and 79.7\% (medical) is explained by systematic differences between participants. This shows that some participants consistently rate explanations higher or lower than others, regardless of the explanation type. The reasons impacting the discrepancy in judgment between participants should be elucidated in future research and considered as an additional predictor. The example accounted for only 6.7\% and 0.9\% of the variance, indicating that the impact of the explanation prototype on evaluation was limited.

\begin{table}[!ht]
\centering
\caption{Results of the mixed effects model analysis for the two datasets}
\label{tab: MEM_results}
\resizebox{\columnwidth}{!}{%
\begin{tabular}{@{}lcccccccccc@{}}
\toprule
 & \multicolumn{5}{c}{\textbf{Telco Dataset}} & \multicolumn{5}{c}{\textbf{Medical Dataset}} \\ \midrule
\textbf{Fixed Effects} & Estimate & SE & 95\% CI & $t$ (df) & $p$ & Estimate & SE & 95\% CI & $t$ (df) & $p$ \\ \midrule
(Intercept) & 29.55 & 1.31 & {[}26.95, 32.14{]} & 22.64 (29.20) & $<$ .001 & 31.16 & 0.99 & {[}29.20, 33.12{]} & 31.33 (61.86) & $<$ .001 \\
Exp. Type {[}Info{]} & $-$1.03 & 0.75 & {[}$-$2.51, 0.45{]} & $-$1.38 (164.41) & .171 & $-$0.08 & 0.45 & {[}$-$0.97, 0.80{]} & $-$0.19 (167.18) & .851 \\ \midrule
\textbf{Random Effects} & \multicolumn{10}{c}{SD (Coefficients)} \\ \midrule
SD (Int: Participant) & \multicolumn{5}{c}{6.99} & \multicolumn{5}{c}{7.01} \\
SD (Int: Example) & \multicolumn{5}{c}{2.43} & \multicolumn{5}{c}{0.74} \\
SD (Residual) & \multicolumn{5}{c}{5.78} & \multicolumn{5}{c}{3.46} \\ \midrule
\textbf{Model Diagnostics} & \multicolumn{10}{c}{ICC} \\ \midrule
Participant & \multicolumn{5}{c}{0.554} & \multicolumn{5}{c}{0.797} \\
Example & \multicolumn{5}{c}{0.067} & \multicolumn{5}{c}{0.009} \\
Adjusted ICC & \multicolumn{5}{c}{0.621} & \multicolumn{5}{c}{0.806} \\ \bottomrule
\end{tabular}%
}
\end{table}

\subsubsection{Comparative Analysis}

Two steps were taken to analyse the forced-choice results. First, the total and per-item number of preferences per dataset was computed. All results are shown in Table \ref{tab: forced_choice}. Subsequently, a mixed-effects logistic regression was implemented to test the statistical significance of the observed trends while accounting for participant and item-wise variation in preference.

\paragraph*{Descriptive Statistics}
Overall, the co-optimised methodology performed better across both datasets ($\Delta = 70$ in the telco dataset and $\Delta = 100$ in the medical dataset). At the question level, the co-optimised explanations performed better on the telco dataset, with only one example (Q6) showing a higher number of preferences for the informative explanation. In the medical dataset, the responses were more mixed, with two examples (Q5 and Q7) showing marginally better performance for the informative explanation. The differences between the two datasets may be linked to their respective domains, with participants likely being less familiar with medical terminology or preferring more technical terms in medical explanations, which may indicate the model's proficiency. This is consistent with the trend observed in the individual evaluation. However, this remains conjecture and requires systematic confirmation. It is important to note that, despite best efforts to counterbalance the question presentation, there is some variation in how many times each example was presented.  

\begin{table}[!ht]
\centering
\caption{Results of the forced-choice analysis with the preferred option for each question in bold.}
\label{tab: forced_choice}
\resizebox{0.8\columnwidth}{!}{%
\begin{tabular}{@{}llllllllllllll@{}}
\toprule
Telco Dataset & Q1 & Q2 & Q3 & Q4 & Q5 & Q6 & Q7 & Q8 & Q9 & Q10 & \textbf{Sum} & \textbf{Mean} & \textbf{SD} \\ \midrule
Informative & 13 & 10 & 9 & 8 & 11 & \textbf{17} & 15 & 11 & 9 & 12 & 115 & 11.5 & 2.84 \\
Co-Optimised & \textbf{18} & \textbf{20} & \textbf{21} & \textbf{23} & \textbf{17} & 12 & \textbf{17} & \textbf{19} & \textbf{19} & \textbf{19} & \textbf{185} & \textbf{18.5} & 2.92 \\ \midrule
Medical Dataset & Q1 & Q2 & Q3 & Q4 & Q5 & Q6 & Q7 & Q8 & Q9 & Q10 & \textbf{Sum} & \textbf{Mean} & \textbf{SD} \\ \midrule
Informative & 5 & 13 & 7 & 4 & \textbf{15} & 12 & \textbf{17} & 9 & 9 & 9 & 100 & 10 & 4.22 \\
Co-Optimised & \textbf{26} & \textbf{18} & \textbf{23} & \textbf{27} & 11 & \textbf{19} & 15 & \textbf{20} & \textbf{20} & \textbf{21} & \textbf{200} & \textbf{20} & 4.78 \\ \bottomrule
\end{tabular}%
}
\end{table}

\paragraph*{Statistical Analysis}
The results of the mixed-effects logistic regression are shown in Table \ref{tab: MEM_comparative}. Choice was specified as the outcome variable, with random intercepts for participant and explanation prototype. The latter was found to have zero variance in the telco dataset and subsequently removed. Overall, the co-optimised explanation was found to be preferred significantly more frequently than expected at chance level, while accounting for participant (and prototype) wise variation. The assumptions of the model were checked using simulation-based scaled residuals, tests of uniformity, outliers and normality of residuals. Influential participants were tested for using leave-one-out refits. All assumptions were found to be satisfied, and tests indicated a good fit.

In the telco dataset, participants selected the co-optimised explanation significantly more often than chance (log-odds $= 0.6$, $SE = 0.20$, $z = 3.05$, $p = 0.002$; $OR = 1.82$, $95\% CI [1.24,2.67]$), which corresponds to an estimated 64.5\% probability of preferring the co-optimised explanation. In the medical dataset, a larger effect in favour of the co-optimised explanations was found (log-odds $= 0.92$, $SE = 0.28$, $z = 3.32$, $p < 0.001$; $OR = 2.52$, $95\% CI [1.46,4.34]$), corresponding to an estimated 71.5\% probability of preferring the co-optimised explanation.

Across the two datasets, participant-level variability was similar ($ICC(telco) = 0.261, ICC(medical) = 0.275$), indicating that individual differences in baseline preference accounted for approximately one quarter of the latent-scale variance for both datasets. Prototype-level variability was small in the medical dataset ($ICC = 0.063$) and zero in the telco dataset, suggesting that the preference for the co-optimised explanations was mostly consistent across the presented instances. 

\begin{table}[!ht]
\centering
\caption{Results of the mixed-effects logistic regression for the comparative evaluation (outcome: co-optimised explanation chosen = 1).}
\label{tab: MEM_comparative}
\resizebox{\columnwidth}{!}{%
\begin{tabular}{@{}lcccccccccc@{}}
\toprule
 & \multicolumn{5}{c}{\textbf{Telco Dataset}} & \multicolumn{5}{c}{\textbf{Medical Dataset}} \\ \midrule
\textbf{Fixed Effects} & Estimate & SE & 95\% CI & $z$ & $p$ & Estimate & SE & 95\% CI & $z$ & $p$ \\ \midrule
(Intercept) & 0.60 & 0.20 & {[}0.22, 1.02{]} & 3.05 & .002 & 0.92 & 0.28 & {[}0.36, 1.55{]} & 3.32 & $<$ .001 \\ \midrule
\textbf{Derived Estimates} & \multicolumn{10}{c}{Back-transformed from the intercept} \\ \midrule
Odds ratio {[}95\% CI{]} & \multicolumn{5}{c}{1.82 {[}1.24, 2.67{]}} & \multicolumn{5}{c}{2.52 {[}1.46, 4.34{]}} \\
$P$(co-optimised) {[}95\% CI{]} & \multicolumn{5}{c}{64.5\% {[}55.3, 72.8{]}} & \multicolumn{5}{c}{71.6\% {[}59.3, 81.3{]}} \\ \midrule
\textbf{Random Effects} & \multicolumn{10}{c}{SD (Coefficients)} \\ \midrule
SD (Int: Participant) & \multicolumn{5}{c}{1.08} & \multicolumn{5}{c}{1.17} \\
SD (Int: Prototype) & \multicolumn{5}{c}{-} & \multicolumn{5}{c}{0.56} \\\midrule
\textbf{Model Diagnostics} & \multicolumn{10}{c}{ICC (latent scale)} \\ \midrule
Participant & \multicolumn{5}{c}{0.261} & \multicolumn{5}{c}{0.275} \\ 
Prototype & \multicolumn{5}{c}{-} & \multicolumn{5}{c}{0.063} \\\midrule
Observations & \multicolumn{5}{c}{300} & \multicolumn{5}{c}{300} \\
Participants & \multicolumn{5}{c}{60} & \multicolumn{5}{c}{60} \\ \bottomrule
\end{tabular}%
}
\begin{tablenotes}
\small
\item Note: The prototype random intercept was omitted from the telco model due to a singular fit (variance estimated at zero).
\end{tablenotes}
\end{table}

Across both evaluation settings, the co-optimised explanation was found to perform better than its informative counterpart. While the individual explanation did not show a significant change in explanation quality, the co-optimised setting yielded directionally higher evaluations. The comparative setting, on the other hand, showed that the co-optimised explanations were selected as more understandable at a level significantly above chance.

\section{Conclusion} \label{sec: Conclusion}

Overall, this study aimed to deploy the FUS to collect US across two datasets, propose and implement an understandability-aware feature selection mechanism and evaluate the quality change of the explanations built on the machine learning model trained on the feature subset. The collected US showed a limited range across both datasets, which may be attributed to a genuine lack of variability in feature understandability but may also be a result of social desirability or a malfunctioning of the scale. Further validation in other domains is necessary to confirm this behaviour. The performance of the random forest remained essentially unchanged between the two settings, while the mean understandability score was successfully increased. While the individual evaluation did not show a significant change in explanation quality, the results were directionally in favour of the co-optimised setting. This was supported by the comparative evaluation, which demonstrated that the co-optimised explanations were considered to be more understandable at a level significantly higher than chance. Given that the forced choice is a more sensitive paired comparison, in contrast to the evaluation scale, which is unvalidated and may be noisier, these results strongly indicate that the integration of understandability into the feature selection process can produce more understandable explanations without degrading model performance.

\section{Limitations and Future Work} \label{sec: Limitations}

The current study took a filter approach to the co-optimisation of understandability and accuracy. While this permitted the maintenance of high accuracy levels and the promotion of understandable features at low computational cost, future research may wish to ascertain whether wrapper or embedded methods can do so more effectively. In addition, the current study treats understandability as a spectrum. However, it may be of interest to discretise the understandability values into 'levels' to ascertain how this affects explanation quality. Furthermore, the current comparative setting requested participants to select explanations they understood better, where future research may wish to measure objective comprehension, rather than subjective understanding.

In addition to implementing other cost-sensitive feature selection methods, future research may consider more complex machine learning models. This study tested a Random Forest for classification, but it may be of interest to assess whether the accuracy-understandability trade-off can be further minimised with more complex models such as Gradient Boosting or small neural networks designed for tabular data, such as TabNet \cite{arik2021tabnet}. While these networks pose some implicit challenges, such as reduced transparency, integrating understandability and employing ad hoc explainability methods may help address them. 

The final explanations were evaluated both individually and comparatively. While the informative explanations were found to perform worse than their co-optimised counterparts, the difference in the individual evaluation was not statistically significant and, especially in the medical dataset, was sufficiently small to potentially be due to sampling noise. While this may reflect a genuine lack of quality difference, the comparative evaluation suggests that the co-optimised explanations are generally understood better, which may be due to the small sample size or overlap in selected features between the two settings. Future research may thus wish to repeat this across different domains and with larger cohorts to confirm trends and assess the extent to which results are reproducible.

Several limitations to the evaluation procedure need to be acknowledged. The current evaluation procedure relied on relatively small sample sizes (236 - 300 observations from 59-60 participants per dataset), limiting the generalisability of the results. Additionally, the scale used to evaluate the quality of explanations has not been formally validated. While the items align with the current study's goal, the validity of the ratings cannot be guaranteed. Based on this, an important avenue for future research may be the psychometric validation of a scale evaluating the quality of explanations. Furthermore, while this study showed some quantitative changes in the quality of the explanations, qualitative interviews may be useful for identifying other areas for improvement in the structure and composition of the explanations, as well as the impact of the quantity and presence of specialised (non-understandable) terms. 

\begin{credits}
\subsubsection{\ackname} This work was conducted with the financial support of Research Ireland - Taighde Éireann, under Grant Nos. 18/CRT/6223 and 12/RC/2289-P2, the latter being co-funded under the European Regional Development Fund. For the purpose of Open Access, the author has applied a CC BY public copyright licence to any Author Accepted Manuscript version arising from this submission.

\subsubsection{\discintname}
The authors have no competing interests to declare that are
relevant to the content of this article.
\end{credits}

\ Appendix
\section*{Appendix}

\renewcommand{\arraystretch}{1.2}
\begin{table}[]
\centering
\caption{Scale adapted from Vilone \& Longo used for the individual explanation evaluation.}
\label{tab: Scale Structure}
\resizebox{\columnwidth}{!}{%
\begin{tabular}{@{}lll@{}}
\toprule
Number & Facet & Item \\ \midrule
1 & Actionability & I have learned something from the explanation \\ 
2 & Causality rev. & The relationship between the input data and the predictions is vague \\ 
3 & Comprehensibility rev. & The structure of the explanation is not clear \\ 
4 & Efficiency rev. & It took me a long time to understand the explanation \\ 
5 & Explicitness rev. & The explanation is incomprehensible \\ 
6 & Informativeness & The explanation provides useful information \\ 
7 & Intelligibility rev. & External support was required to understand the explanation e.g., books, internet search, another person \\
8 & Interestingness rev. & The explanation is not interesting \\
9 & Mental fit & The explanation allows me to understand the logic of the machine learning model used to generate the predictions \\ 
10 & Security & Thanks to the explanation, I believe that the model will return accurate predictions for all reasonable inputs \\ \bottomrule
\end{tabular}%
}
\end{table}

\begin{table}[!ht]
\centering
\caption{``Please rate the feature `Credit Score' using the following questions. The scores of `Credit Score' range from 18.4 to 118.65. The feature is used to predict whether or not a customer's loan application will be approved. While rating the feature, consider if you would understand a loan decision, if it was justified using the feature. E.g. 'The customer did not get a loan because their Credit Score was 56.95'."}
\label{tab: example}
\resizebox{\columnwidth}{!}{%
\begin{tabular}{lccccc}
\hline
 & \multicolumn{1}{l}{Strongly Disagree} & \multicolumn{1}{l}{Disagree} & \multicolumn{1}{l}{Neutral} & \multicolumn{1}{l}{Agree} & \multicolumn{1}{l}{Strongly Agree} \\ \hline
I can understand the scale (units) of the feature. & x &  &  &  &  \\
I can easily understand if a given value of the feature is high or low. &  &  &  & x &  \\
I know what this feature measures. &  &  &  & x &  \\
I know what this feature represents. &  & x &  &  &  \\
I think that I can easily access a definition of the feature. &  &  &  &  & x \\
In my opinion, the feature should be used to predict the outcome. &  &  & x &  &  \\
I think that the feature is important for the outcome. &  &  &  &  & x \\
I think it is fair that the feature influences the outcome. &  &  &  & x &  \\ \hline
\end{tabular}%
}
\end{table}


\bibliography{references}

\end{document}